\documentclass[twocolumn,showpacs,prx,preprintnumbers,amsmath,amssymb,floatfix,eqsecnum]{revtex4}
\usepackage{graphicx}
\usepackage{epsfig,psfrag}
\usepackage{dcolumn}
\usepackage{bm}
\usepackage{color}
\begin{document}
\title{ Quasiparticle trapping, Andreev level population dynamics, and 
charge imbalance in superconducting weak links}

\author{A.~Zazunov,$^1$ A.~Brunetti,$^1$ A.~Levy Yeyati,$^2$ and R.~Egger$^1$}

\affiliation{${}^1$~Institut f\"ur Theoretische Physik,
Heinrich-Heine-Universit\"at, D-40225  D\"usseldorf, Germany\\
${}^2$~Departamento de F{\'i}sica Te{\'o}rica de la
Materia Condensada C-V and Instituto Nicol\'as Cabrera,
Universidad Aut{\'o}noma de Madrid, E-28049 Madrid, Spain }

\date{\today}
\begin{abstract} 
We present a comprehensive theoretical framework for the Andreev bound 
state population dynamics in superconducting weak links. 
Contrary to previous  works, our approach takes
into account the generated nonequilibrium distribution of the 
continuum quasiparticle states in a self-consistent way.  
As application of our theory, we show that the coupling 
of the superconducting contact to environmental phase fluctuations 
induces a charge imbalance of the continuum quasiparticle population. 
This imbalance is due to the breaking of the left-right symmetry in the rates 
connecting continuum quasiparticles and the Andreev bound state system, and
causes a quasiparticle current on top of the Josephson current in a ring geometry.  
We evaluate the phase dependence
of the quasiparticle current for realistic choices of the model parameters.
Our theory also allows one to analyze the quantum coherent evolution of the system from 
an arbitrary initial state.
\end{abstract}
\maketitle

\section{Introduction}\label{sec1}

Quantum coherent superconducting circuits are among the most promising
candidates for future large-scale quantum information processing devices,
and the last few years have seen an enormous increase in research activity
in this direction \cite{devoret,nazarov}.  
Their functioning is often limited by the presence of residual 
nonequilibrium quasiparticles, 
whose uncontrolled tunneling provide a severe decoherence mechanism 
\cite{poison1,poison2,poison3,poison4}. Remarkably, 
in some cases where the parity of the quantum state matters, 
the presence of a single extra quasiparticle can determine the 
macroscopic response of the device \cite{poison5,poison6}.  
The trapping of single quasiparticles in superconducting islands is known as
``poisoning" \cite{aumentado}.
Although at temperatures well below the superconducting gap $\Delta$,
such states have an exponentially small chance to exist in thermal
equilibrium, they can have very long lifetimes once they are generated
in a nonequilibrium process.  Quasiparticle poisoning was also 
observed in  recent experiments \cite{zgirski,bretheau1,bretheau2} 
for devices containing a short superconducting weak link with only a few 
transport channels. Those experiments reported the existence of 
long-lived nonequilibrium quasiparticles trapped in the Andreev bound 
states \cite{alfredo} formed near the weak link.  
Here, we refer to such 
a superconducting constriction as a ``superconducting atomic contact'' 
(SAC) \cite{agrait}.

In this paper, we provide a comprehensive theoretical framework
for the understanding of the Andreev bound state population dynamics 
in a single-channel SAC.  In our theory, 
transitions between different Andreev bound state configurations and 
their interplay with continuum quasiparticles  
are fully taken into account. Such transitions can be triggered, for instance,
by environmental phase fluctuations or by phonon-induced processes, 
and also allow for a change of the fermion number parity
in the Andreev levels. While the ground state has even parity, 
there are two spin-degenerate odd-parity Andreev bound state configurations
 with excitation energy 
\begin{equation}\label{absen}
 E_A(\varphi_0)= \Delta \sqrt{ 1-{\cal T}\sin^2(\varphi_0/2) }
\end{equation}
relative to the ground state, where $\varphi_0$ is the superconducting
phase difference across the contact and ${\cal T}$ the 
normal-state transmission probability of the contact.
The occupation of such an odd-parity state causes  
quasiparticle poisoning since the Josephson current is 
then completely blocked.  Due to its long lifetime, the spin degree of 
freedom corresponding to the two odd-parity states
has also been proposed as qubit platform \cite{chtech,padurariu}.
On the other hand, the occupation of odd-parity states 
severely limits the operation of the ``Andreev qubit'' 
\cite{zazunov1,zazunov2}, which is built from
the Andreev ground state configuration and the excited even-parity  
state of energy $2E_A$, cf.~also Ref.~\cite{desposito};
for ${\cal T}\to 1$ and $\varphi_0\approx \pi$, these two states
are nearly degenerate.
Similar superconducting devices are also discussed in the 
context of Majorana fermion physics \cite{heck,rainis},
and questions pertaining to quasiparticle poisoning and the 
interplay between Andreev (or Majorana) and continuum 
quasiparticle distribution functions are important in that direction as well.
The phase-dependent energy $E_A$ in Eq.~\eqref{absen} also determines the 
transition frequencies between different Andreev configurations, which
have recently been studied by microwave absorption and 
supercurrent spectroscopy \cite{bretheau1,bretheau2}, where the 
odd-parity states can be excited together with a continuum quasiparticle. For the
theoretical description of such ``artifical atom''  spectra 
in microwave irradiated contacts, see
Refs.~\cite{desposito,kos,bretheau3} and references therein.

\begin{figure}
\begin{center}
\includegraphics[width=0.48\textwidth]{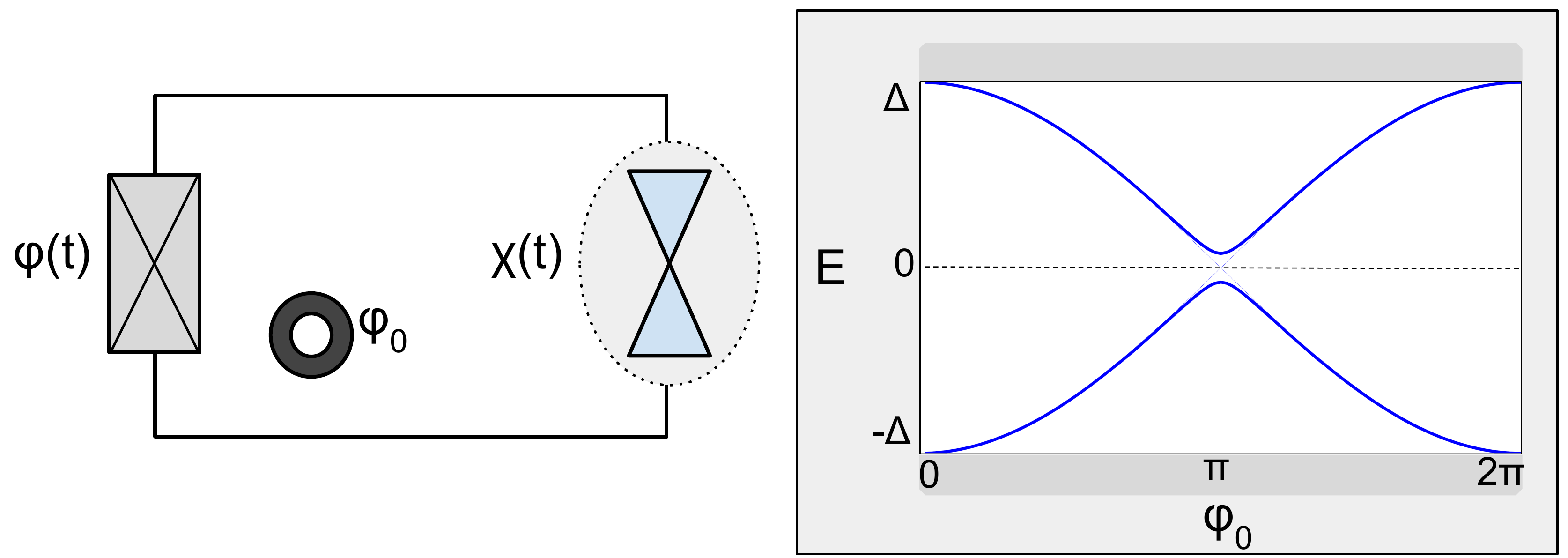}
\end{center} 
\caption{\label{fig1} 
Schematic illustration of the SQUID geometry considered in this work,
where the SAC is embedded into a ring containing a conventional
Josephson junction that generates environmental phase fluctuations.
The Andreev bound states are depicted in the right panel. 
For details, see text. 
}
\end{figure}

The above discussion shows that it is an important issue to 
 understand the Andreev bound state population dynamics 
and its interplay with continuum quasiparticles 
in SACs. We here study this problem for the
simplest single-channel case, where the SAC quasiparticles 
interact with the phase fluctuations of 
an electromagnetic environment.
To be concrete, we shall consider the plasma mode of another
Josephson junction in the ring geometry of Fig.~\ref{fig1},
but our subsequent discussion is formulated for a general
environment.  Previous work has already addressed some
aspects of this problem \cite{olivares}, but to the best
of our knowledge, the generated nonequilibrium distribution
of continuum quasiparticles and its interplay with quasiparticle
poisoning has not been discussed in a self-consistent manner up to now.

The structure of the remainder of this paper is as follows.
In Sec.~\ref{sec2}, we introduce the model and provide a 
second-quantized formulation, where the
fermionic (Andreev bound state and continuum) quasiparticles 
are weakly coupled to the environmental phase fluctuations.
In Sec.~\ref{sec3}, assuming that the electromagnetic environment
remains in thermal equilibrium, we provide the 
master equation description of this system. 
The density matrix for the quasiparticles can then be factorized 
into an Andreev part, $\rho_A(t)$, and a diagonal density matrix 
describing the quasiparticles belonging to the continuum spectrum. 
Importantly, the off-diagonal part of $\rho_A(t)$ decouples from 
the diagonal part, where the latter is determined by 
Andreev level occupation probabilities.  Including quasiparticle
relaxation by phonons in a phenomenological way, the 
resulting steady-state solution can be obtained by self-consistently
solving just two coupled nonlinear equations.  
In Sec.~\ref{sec4}, as an application of our approach, we
describe a charge imbalance effect caused by an 
asymmetry in the transition rates  between 
Andreev and continuum quasiparticles.  The self-generated nonequilibrium
distribution of continuum quasiparticles (we note that no 
external forces drive  the system out of equilibrium in our model)
causes a phase-dependent quasiparticle current, and
an asymmetric charge profile around the weak link.
The predicted charge imbalance effect could be measured by  
superconducting-normal tunnel junction spectroscopy 
\cite{snj1,snj2,snj3,snj4}.   Finally, we conclude in Sec.~\ref{sec5}.
Details about our calculations can be found in several Appendices.
We often use units with $\hbar=k_B=e=1$.

\section{Theoretical framework}\label{sec2}

\subsection{Model}

In this work, we consider a single-channel SAC embedded in the asymmetric 
SQUID geometry sketched in Fig.~\ref{fig1}, where the ring contains 
both the SAC and a conventional Josephson junction.   
This setup has also been studied in other theoretical works, for instance,
see Refs.~\cite{olivares,bretheau3}.
Denoting the superconducting phase differences across
the SAC and the Josephson junction by $\varphi$ and $\chi$, respectively,
both phases are linked by
\begin{equation}\label{link}
\chi(t)-\varphi(t)+\varphi_0 = 0,
\end{equation} 
where the dimensionless parameter $\varphi_0$ is related to the magnetic flux 
(in units of the flux quantum $h/2e$) threading the ring.  
Assuming that the Josephson energy, $E_J$, is 
much bigger than the charging energy, $E_C=(2e)^2/2C$ with the
capacitance $C$, of the 
Josephson junction, the environmental electromagnetic modes 
seen by the SAC are well described by an effective $LC$ circuit Hamiltonian,
\begin{equation}\label{Henv1}
H_{\rm env} = - E_C \frac{d^2}{d\chi^2} + \frac{E_J}{2} \chi^2,
\end{equation}
corresponding to an undamped harmonic oscillator.
In Sec.~\ref{sec3}, we shall also include the effects of an additional
shunt resistance $R$, leading to the damping parameter 
$\eta_d=1/(RC)$.
Note that in the regime $E_J\gg E_C$ of interest here, 
fluctuations of $\chi$ are small, $\langle \chi^2\rangle\ll 1$. 
Turning to the single-channel SAC, the BCS Hamiltonian 
is written in terms of a two-component Nambu
 spinor, $\psi(x)= (\psi_\uparrow(x),\psi^*_\downarrow(x))^T$,
describing electrons in the left ($x<0$) or right ($x>0$) 
superconducting bank, with the contact at $x=0$.
Using the standard quasiclassical Andreev approximation \cite{nazarov}, 
we introduce slowly varying envelope functions, 
$\psi(x)=e^{ik_F x} \psi_{R}(x)+e^{-ik_F x} \psi_L(x)$, 
with Fermi momentum $k_F$.  
Combining the right- and left-moving envelopes into 
$\Psi(x)=(\psi_R,\psi_L)^T$, where each entry still carries 
the Nambu spinor structure, the time-dependent 
wave function satisfies the Bogoliubov-de Gennes (BdG) equation 
\cite{nazarov}, 
\begin{eqnarray}\label{bdg}
&& \left( i\partial_t - H_{\rm BdG} \right) \Psi(x,t) = 0,\\
&& H_{\rm BdG} = - i v_F \tau_z\sigma_z\partial_x + \Delta\tau_0\sigma_x,
\nonumber
\end{eqnarray} 
with Fermi velocity $v_F$, the BCS gap $\Delta$, and
Pauli matrices $\sigma_{x,y,z}$ and $\tau_{x,y,z}$ in 
Nambu and right/left-mover space, respectively; the corresponding unit 
matrices $\sigma_0$ and $\tau_0$ are often kept implicit below. As 
shown in Ref.~\cite{zazunov2} and references therein, 
the BdG solutions on both sides of the contact have 
to be matched at $x=0$ by a transfer matrix, 
\begin{equation}\label{matching}
\Psi(-0^+,t) = \frac{e^{i\sigma_z\varphi(t)/2}} {\sqrt{\cal T}}
\left( \tau_0 + \sqrt{1-{\cal T}} \tau_x \right)\Psi(0^+,t),
\end{equation}
which is $4\pi$-periodic in  $\varphi$. 
For simplicity, the transmission probability, 
$0<{\cal T}\le 1$, which characterizes the transparency of the 
constriction in the normal phase, is assumed
energy-independent. 
In what follows, it is convenient to remove 
the time dependence from Eq.~\eqref{matching} by a gauge transformation,
\begin{equation}\label{gauge}
\Psi(x,t)\to e^{- (i/4) {\rm sgn}(x) \chi(t) \tau_0\sigma_z} \Psi(x,t),
\end{equation}
with $\chi(t)$ in Eq.~\eqref{link}. 
The phase factor in Eq.~\eqref{matching}
thereby becomes time-independent, with $\varphi(t)\to \varphi_0$,
and $H_{\rm BdG} \to H_{\rm BdG} + V$, where the interaction term
is, with $\dot\chi=\partial_t \chi$, given by
\begin{eqnarray}\label{vint}
V(x,t) &=& A(x) 
\dot\chi(t) + W(x) \chi(t) + {\cal O}\left(\chi^2\right), \\ \nonumber
A(x) &=& -\frac{1}{4} {\rm sgn}(x)\tau_0 \sigma_z,  \quad
W(x) = -\frac{\Delta}{2} {\rm sgn}(x) \tau_0\sigma_y.
\end{eqnarray}
Since $\langle\chi^2\rangle\ll 1$, 
the linearized expression in Eq.~\eqref{vint} is sufficient, which now 
couples the quasiparticle dynamics to the phase $\chi(t)$.
Using the Josephson plasma frequency,
$\Omega=\sqrt{2E_C E_J}$, where we assume $\Omega<\Delta$
throughout this paper, the Lagrangian of the coupled system,
with $\bar\Psi=(\Psi^*)^T$, is
\begin{equation}
L(t) = \frac{1}{4E_C} \left(\dot\chi^2 -\Omega^2\chi^2 \right)  
+ \int dx \ \bar\Psi ( i\partial_t - H_{\rm BdG}-V)\Psi.
\end{equation}
Employing the momentum $P_\chi$ canonically conjugate to the phase 
 $\chi$, the corresponding Hamiltonian is
\begin{eqnarray} \nonumber
H&=& E_C\left( P_\chi + \int dx\  \bar \Psi A(x) \Psi \right)^2+
\frac{\Omega^2}{4E_C} \chi^2 
\\ \label{ham1}
&+& \int dx\ \bar \Psi [H_{\rm BdG}+W(x)\chi] \Psi.
\end{eqnarray}

\subsection{Second-quantized formulation}

We now switch to a second-quantized language by letting
 $\Psi(x)\to \hat \Psi(x)$, where the electron field operator, $\hat \Psi(x)$,
is expanded in terms of the 
stationary solutions, $\Psi_\nu(x)$, with energy $E_\nu$, 
of the BdG equation for  
time-independent matching condition \eqref{matching},  
i.e., for $\varphi(t)=\varphi_0$. 
The wave functions $\Psi_\nu(x)$ thus represent the noninteracting 
SAC eigenstates. 
Introducing the corresponding quasiparticle creation (annihilation) operators
$\gamma^\dagger_\nu$ ($\gamma_\nu^{}$), with the 
standard fermionic anticommutator algebra  
$\{ \gamma_\nu^{} , \gamma^\dagger_{\nu'} \}=\delta_{\nu\nu'}$, we arrive
at $\hat \Psi(x)=\sum_\nu \Psi_\nu(x)\gamma_\nu^{}$. The
noninteracting SAC Hamiltonian then reads
\begin{equation}\label{sac}
H_{\rm SAC} = \sum_\nu E_\nu\gamma_\nu^\dagger \gamma_\nu^{}.
\end{equation} 
The quantum numbers $\nu$ include (i) a pair of Andreev bound states,
$\nu=\eta=\pm$, where the energy $E_\eta=\eta E_A$, with $E_A(\varphi_0)$
in Eq.~\eqref{absen}, is within the BCS gap
and $\Psi_\eta(x)$ stays localized near the contact at $x=0$, 
and (ii) delocalized scattering states in the continuum,
$\nu=p=(E,s)$, where $|E|\ge  \Delta$ and the 
index $s$ (with $s=1,2,3,4$) refers to the four possible types of 
incoming states (from the left or right side, and of electron- or 
hole-like character).  The wave functions $\Psi_\nu(x)$ 
are provided in analytical form in Appendix~\ref{appa}, see also
Ref.~\cite{olivares}.

We here employ a semiconductor representation,
where quasiparticles are effectively spinless but can have either 
positive or negative energy.  In the ground state of $H_{\rm SAC}$, 
all $E_\nu<0$ states are occupied, including the $\eta=-$ Andreev 
bound state.  Using standard occupation 
number operators, $\hat n_\nu=\gamma_\nu^\dagger \gamma_\nu^{}$, with
 eigenvalues $n_\nu=0,1$, the four possibilities for the 
occupation of the Andreev bound state sector are indexed by 
$(n_{+}, n_{-})$.  The ground state, with energy $-E_A$, corresponds
to the $(0,1)$ configuration, 
which we also denote by the Andreev state
 $|-\rangle_A$.  This state carries the equilibrium 
Josephson supercurrent $I_A=-(2e/\hbar) \partial E_A/\partial \varphi_0$. 
In the parlance of Refs.~\cite{zgirski,olivares},
$|-\rangle_A$ is an even-parity state, while 
the odd-parity sector corresponds to the spin-degenerate 
$(0,0)$ and $(1,1)$ states, with excitation energy  $E_A$ relative 
to the ground state.
The odd-parity states with $n_++n_-=0$ and $2$, resp., are denoted by
\begin{equation}\label{oddpar}
|0\rangle_A= \gamma_- |-\rangle_A, 
\quad |2\rangle_A= \gamma_+^\dagger|-\rangle_A,
\end{equation}
and imply a vanishing Andreev supercurrent, consistent with the
``quasiparticle poisoning'' scenario.
The lifetime of these states can reach the millisecond regime for 
high transparency ${\cal T}$,  
and their decay rate exhibits nearly universal scaling 
as a function of $E_A/\Delta$ \cite{zgirski,olivares}.
Finally, the $(1,0)$ even-parity state, denoted  as
\begin{equation}\label{evenpar}
|+\rangle_A=\gamma_+^\dagger\gamma_-|-\rangle_A,
\end{equation} 
represents an excited ``Andreev Cooper pair'' 
localized at the contact, with 
excitation energy $2E_A$ above the ground state.  
 The $|+\rangle_A$ state carries the
Josephson current $-I_A$, with opposite sign as compared to $|-\rangle_A$, 
but rather quickly relaxes to the ground state \cite{olivares}.

The second-quantized form of the interacting
Hamiltonian (\ref{ham1}) is thus given by
\begin{equation}
H = E_C (P_\chi+\bar A)^2 +\frac{\Omega^2}{4E_C}\chi^2 
+ H_{\rm SAC} + \chi \sum_{\nu,\nu^\prime} 
W_{\nu\nu^\prime}\gamma_\nu^{\dagger} \gamma_{\nu^\prime}^{}, 
\end{equation}
where $\bar A  =  \sum_{\nu,\nu^\prime} 
A_{\nu\nu^\prime} \gamma^\dagger_{\nu} \gamma_{\nu^\prime}^{}$ 
plays the role of a vector potential.  The matrix elements 
\begin{equation}\label{awmat}
A_{\nu\nu^\prime} = \int dx \ 
\Psi^\dagger_\nu  A(x) \Psi_{\nu^\prime} , \quad
W_{\nu\nu^\prime}= \int dx \ \Psi_\nu^\dagger W(x) 
\Psi_{\nu^\prime}^{},
\end{equation}
are discussed below and in App.~\ref{appb}.
For convenience, we now shift $P_\chi\to P_\chi-\bar A$ by means of 
a unitary transformation, $H\to UHU^\dagger$ with $U=e^{i\bar A\chi}$,
and represent the (unitarily transformed) phase $\chi$ and 
its momentum $P_\chi$ by a standard boson operator, $b$, with commutator
$[b,b^\dagger]=1$, such that $\chi=\sqrt{E_C/\Omega}\ (b+b^\dagger)$. 
After some algebra, we thereby arrive at the Hamiltonian in 
its final form (up to an irrelevant constant),
\begin{equation}\label{finalham}
H = H_{\rm SAC}+  \Omega b^\dagger b 
+ \lambda\left (b+b^\dagger\right) \hat I_S,
\end{equation}
describing fermionic (Andreev level and continuum) 
quasiparticles coupled to an oscillator mode with the plasma 
frequency $\Omega$.   For $E_J\gg E_C$, we are effectively in the 
weak-coupling regime, $\lambda\ll 1$, with the dimensionless 
coupling strength $\lambda= \sqrt{E_C/4\Omega}$.
Finally, the Josephson current operator in Eq.~\eqref{finalham} is 
\begin{eqnarray}\nonumber
\hat I_S &=& \sum_{\nu,\nu^\prime} {\cal I}_{\nu\nu^\prime} 
\gamma_\nu^\dagger \gamma_{\nu^\prime}^{},\\ 
\label{joscur}
{\cal I}_{\nu\nu^\prime} &=& 2 W_{\nu\nu^\prime} -
 2 i \left(E_\nu-E_{\nu^\prime} \right) A_{\nu\nu^\prime},
\end{eqnarray}
where the matrix elements 
${\cal I}_{\nu\nu^\prime}$ are  discussed in App.~\ref{appb} and in the 
next subsection.

\subsection{Current}

We first note that due to the spatial homogeneity of the extended
quasiparticle states (plane waves) away from the contact, 
the matrix elements $A_{pp'}$ and $W_{pp'}$, and hence also ${\cal I}_{pp'}$,
between continuum states can be finite only when their
energies match, $E=E'$, i.e.,
phase fluctuations do not induce intraband transitions.
As we show in App.~\ref{appb}, one finds that
$W_{pp'}=0$ even for $E=E'$, implying that ${\cal I}_{pp'}=0$.
Therefore, delocalized continuum states 
can contribute to the supercurrent $\hat I_S$ only 
through transitions mixing
them with Andreev levels. 

The Josephson current operator then contains a 
part $\hat I_A$, coming from the Andreev sector only,
and a part $\hat I_{cA}$, describing the mixing of
continuum and Andreev states,  $\hat I_S= \hat I_A+\hat I_{cA}$. 
We find for the pure Andreev current \cite{zazunov1,zazunov2}
\begin{eqnarray}\label{iacur}
\hat I_A &=& - \frac{{\cal T}\Delta^2 \sin(\varphi_0/2)}{E_A}
\\ \nonumber &\times& \gamma^\dagger 
\left[ \cos(\varphi_0/2) \eta_z
-\sqrt{1-{\cal T}} \sin(\varphi_0/2) \eta_y \right] \gamma,
\end{eqnarray}
where $\gamma=(\gamma_+,\gamma_-)^T$ combines the two Andreev 
level fermion operators, and the 
Pauli matrices $\eta_{x,y,z}$ act in the corresponding space.
Note that the Andreev current operator \eqref{iacur} is written in 
energy representation, where the Hamiltonian projected to the 
Andreev sector is diagonal, $H_A= E_A \gamma^\dagger \eta_z \gamma$.
For non-ideal transparency of the contact, ${\cal T}<1$, 
$\hat I_A$ does not commute with $H_A$ ---
 Andreev level eigenstates are
superpositions of current eigenstates implying that strong
fluctuations of the supercurrent are
generated for $\varphi_0\approx \pi$ \cite{martin-rodero}.

Similarly, the supercurrent contribution caused by the mixing of
continuum and Andreev states is 
\begin{equation} \label{ica}
\hat I_{cA}=\sum_{\eta=\pm} \sum_{p=(E,s)} {\cal I}_{\eta,p} 
\gamma_\eta^\dagger \gamma_p^{}  + {\rm h.c.},
\end{equation}
where the matrix elements ${\cal I}_{\eta,p}$ are specified
in App.~\ref{appb}.

Finally, the total current flowing through the contact
also contains a conventional dissipative quasiparticle contribution
due to continuum states, $I_{\rm qp}$, on top of the supercurrent contribution
$\langle \hat I_S\rangle$.  We provide the 
standard scattering theory expression for $I_{\rm qp}$ in App.~\ref{appb}.

To study the physics described by the interacting Hamiltonian, $H=H_0+V$,
with the noninteracting piece 
$H_0=H_{\rm SAC}+\Omega b^\dagger b$ and the interaction contribution
$V= \lambda(b+b^\dagger)\hat I_S$, we now turn to a master equation 
approach.  In this work, we assume that the plasma mode remains in
thermal equilibrium with a heat bath of temperature $T_{\rm env}$ at all times, 
and we thus neglect feedback effects on the phase dynamics.  

\section{Master equation approach}\label{sec3}

\subsection{Master equation}

Within the master equation framework \cite{nazarov}, 
the Liouville-von Neumann equation for the density matrix of 
the complete system, $\rho_{\rm tot}$, 
is expanded to second order in the small interaction 
parameter $\lambda\ll 1$.  
Writing time-dependent operators in the interaction 
picture as ${\cal O}(t)=e^{iH_0 t} {\cal O}
e^{-i H_0 t}$, the density matrix then obeys the equation
($[A,B]$ denotes the commutator)
\begin{eqnarray}\nonumber
\partial_t \rho_{\rm tot}(t) &=& -
\int_0^t d\tau \ [V(t),[V(t-\tau),\rho_{\rm tot} (t-\tau)] ] \\
&-& i [V(t),\rho_{\rm tot}(0)].
\label{liou1}
\end{eqnarray}
Our assumption of thermal equilibrium for the plasma mode implies
a factorized form of the density matrix,
\begin{equation}
\rho_{\rm tot}(t)= \rho_{\rm osc}\otimes \rho(t),
\end{equation}
where $\rho_{\rm osc}\sim e^{-(\Omega/T_{\rm env}) b^\dagger b}$ is a
thermal density matrix for the plasma mode 
and $\rho(t)$ describes the time evolution of fermionic quasiparticles.   
Taking the trace over the oscillator degree of freedom, 
Eq.~\eqref{liou1} yields
\begin{eqnarray} \label{liou2}
\partial_t \rho(t) &=& -\int_0^\infty d\tau
\Bigl[ D(\tau) \hat I_S(t) \hat I_S(t-\tau) \rho(t) \\ \nonumber
&-& D(-\tau)\hat I_S(t)\rho(t)\hat I_S(t-\tau) \Bigr] + {\rm h.c.},
\end{eqnarray}
where we have employed the Markov approximation, valid 
at long times $t$ and not too low temperatures \cite{nazarov,foot1}.
The boson correlator in Eq.~\eqref{liou2} reads
\begin{equation}\label{dome}
D(\tau) = \int_0^\infty d\omega J(\omega) \left[
\left(n_B(\omega)+1\right)e^{-i\omega\tau}+n_B(\omega) e^{i\omega\tau}\right],
\end{equation}
with the Bose function,  
\begin{equation}\label{bose}
n_B(\omega)=\frac{1}{e^{\omega/T_{\rm env}}-1},
\end{equation}
and the environmental spectral density,
\begin{equation}\label{spectral}
J(\omega)= \frac{\lambda^2\eta_d}{2\pi} \left(
\frac{1}{(\omega-\Omega)^2+\eta_d^2/4}-
\frac{1}{(\omega+\Omega)^2+\eta_d^2/4}\right).
\end{equation}
We use Eq.~\eqref{spectral} below also for $\omega<0$, 
and directly include the Ohmic damping parameter, $\eta_d$, 
to capture the effects of a shunt resistance, see Sec.~\ref{sec2}. 
For $\eta_d\to 0$, the spectral density has the limit $J(\omega)
= \lambda^2 \delta(|\omega|-\Omega) {\rm sgn}(\omega)$.
For finite $\eta_d$, Eq.~\eqref{spectral} exhibits sharp peaks for 
$|\omega|=\Omega$. 

The equation of motion (\ref{liou2}) is still quite 
cumbersome, and we shall here proceed by making two approximations.
First, we neglect entanglement between the Andreev and 
continuum quasiparticles, which means that the reduced density matrix 
factorizes into an Andreev part and a continuum part, 
\begin{equation} \label{factoriz}
\rho(t)=\rho_A(t) \otimes \rho_c(t).
\end{equation}
This approximation is justified in the weak-coupling regime $\lambda\ll 1$,
since higher-order terms in $\lambda$ are needed to coherently couple
Andreev and continuum states \cite{foot2}.
The factorized density matrix \eqref{factoriz} is expected to be 
highly accurate away from the zero-temperature limit, 
since the thermal energy uncertainty causes a blurring of 
continuum quasiparticle wavepackets that  
rapidly destroys entanglement between Andreev and continuum states.  
Second, we also assume that the density matrix $\rho_c(t)$ describing
continuum quasiparticles remains diagonal during the time evolution.
This approximation is justified by noting that there are no direct matrix 
elements in $H$ connecting different continuum states, 
and implies that $\rho_c(t)$ is fully determined by specifying
the time-dependent occupation probabilities $n_p(t)$ of continuum
states, 
\begin{equation}\label{denscon}  
\rho_c(t)=\prod_{p} \Bigl[ n_p(t) \left|1_p\right\rangle\left\langle 1_p\right|
+[1-n_p(t)] \left|0_p\right\rangle\left\langle 0_p\right| \Bigr],
\end{equation}
where $|1_p\rangle= \gamma_p^\dagger|0_p\rangle$ corresponds to a 
filled single-particle state $p=(E,s)$. 
Note that $\rho_c(t)$ in Eq.~\eqref{denscon} is always normalized,
${\rm Tr}_c\left[\rho_c(t)\right]=1$.  On the other hand, the density 
matrix $\rho_A(t)$ describing the Andreev sector,
 with normalization condition ${\rm Tr}_A\left[\rho_A(t)\right]=1$,
may have off-diagonal entries reflecting quantum coherence.

Tracing over the Andreev part in Eq.~\eqref{liou2} then yields an 
equation of motion for the continuum state occupation numbers $n_p(t)$.
Similarly, tracing instead over the continuum states, one obtains
an equation for the time evolution of the reduced Andreev density matrix
$\rho_A(t)$.  In these equations, the transition rates between
different levels follow as Fermi golden rule expressions,
\begin{equation}\label{rate}
\Gamma_{\nu\nu'} = \frac{2\pi}{\hbar} \left| {\cal I}_{\nu\nu'} \right|^2  
\left[1+n_B\left(E_\nu-E_{\nu'}\right)\right] J\left(E_\nu-E_{\nu'}\right),
\end{equation}
with the Bose function $n_B(\omega)$ in Eq.~\eqref{bose} and the
spectral density $J(\omega)$ in Eq.~\eqref{spectral}.
By using Eq.~\eqref{iacur}, we observe that 
the direct rates connecting different Andreev states are given by
\begin{eqnarray}\label{directrate}
\Gamma_{\eta,-\eta} &=& \frac{2\pi}{\hbar} (1-{\cal T})
\frac{(\Delta^2-E_A^2)^2}{E_A^2} \\ \nonumber &\times&
\left[\delta_{\eta,+}+ n_B(2 E_A) \right] J(2 E_A).
\end{eqnarray}
Notice that these rates vanish for perfect transparency, ${\cal T}\to 1$.
Recalling now that ${\cal I}_{pp'}=0$ for arbitrary ${\cal T}$, 
we see that transition rates between continuum states 
are always absent, $\Gamma_{pp'}=0$.
Finally, the supercurrent matrix elements between Andreev and continuum states,
${\cal I}_{\eta p}$, see Eq.~(\ref{joscur}) and
App.~\ref{appb}, determine the corresponding transition rates, 
$\Gamma_{\eta,p}$, for exciting an Andreev quasiparticle into the 
continuum, plus the reverse process with rate $\Gamma_{p,\eta}$.   
Such transitions must involve the absorption or emission
of an environmental photon. Since $|E|\ge \Delta$ and 
the spectral density is sharply peaked
around the Josephson plasma frequency $\Omega$, those rates are sizeable
only when $\Omega> \Delta-E_A$ \cite{olivares}.

Performing now the trace over the Andreev sector in Eq.~\eqref{liou2}, we find
\begin{equation} \label{liou3a}
  \partial_t n_p=  - \sum_{\eta=\pm} \left[ 
\Gamma_{p,\eta} (1-n_\eta) n_p - \Gamma_{\eta,p} (1-n_p) n_\eta\right].
\end{equation}
The time-dependent continuum state distribution function, $\{n_p(t)\}$,
thereby couples to the Andreev level  occupation probabilities,
\begin{equation}\label{nplusminus}
n_\eta(t)= {\rm Tr}_A \left[ \hat n_\eta \rho_A(t) 
\right] , \quad \hat n_\eta=
\gamma_\eta^\dagger \gamma_\eta^{} .
\end{equation}
Tracing instead over the continuum states in Eq.~\eqref{liou2},
we find ($\{A,B\}$ denotes the anticommutator)
\begin{eqnarray} \nonumber
&& \partial_t\rho_A(t)= -\frac12 \sum_\eta  
 \Gamma_{\eta,-\eta} \left \{ 
\hat n_\eta(1-\hat n_{-\eta}) , \rho_A(t) \right \} \\ \label{rhoAeq}
 &&+ \sum_\eta \Gamma_{-\eta,\eta} \gamma_\eta^\dagger\gamma_{-\eta}^{}
\rho_A(t) \gamma_{-\eta}^\dagger \gamma_\eta^{} \\ 
\nonumber && - \sum_{p,\eta} \Gamma_{p,\eta} n_p(t) \left(
\frac12 \left \{1-\hat n_\eta,\rho_A(t) \right\} -
\gamma_\eta^\dagger \rho_A(t) \gamma_\eta^{} \right) \\
\nonumber &&  - \sum_{p,\eta} \Gamma_{\eta,p}[1-n_p(t)] 
\left (\frac12\left\{ \hat n_\eta,\rho_A(t)\right\} -\gamma_\eta^{}
\rho_A(t) \gamma_\eta^\dagger \right) .
\end{eqnarray}
This equation has been obtained under the assumption that the coupling to the plasma
mode provides the only relaxation mechanism, but in Eq.~\eqref{steady1} below, we will also 
include the effect of other mechanisms (e.g., phonons) through a phenomenological
relaxation time $\tau_{\rm qp}$. 
Notice that the terms $\sim \gamma_\eta^\dagger \rho_A(t) \gamma^{}_\eta$
and $\sim \gamma_\eta^{}\rho_A(t)\gamma^\dagger_\eta$ in Eq.~\eqref{rhoAeq}
describe ``parity jumps'' where the fermion number parity 
of Andreev quasiparticles can change.

Since there are four Andreev configurations ($n_+,n_-$), 
the Andreev density matrix is a $4\times 4$ matrix.  
We here represent $\rho_A(t)$ in the basis spanned by the 
Andreev ground state $|-\rangle_A$,
corresponding to the $(0,1)$ configuration, the spin-degenerate
odd-parity states $|0\rangle_A$ and $|2\rangle_A$ in 
Eq.~\eqref{oddpar}, and the excited 
even-parity state $|+\rangle_A$ in Eq.~\eqref{evenpar}.
The diagonal elements of $\rho_A(t)$ yield
the respective occupation probabilities, 
$P_0(t)={}_A\langle 0|\rho_A(t)|0\rangle_A$, and likewise
for $P_{\eta=\pm}(t)$ and $P_2(t)$.  Thereby the normalization 
condition for $\rho_A(t)$ gives 
\begin{equation}\label{normP}
P_0(t) + P_2(t) + \sum_{\eta} P_\eta(t) = 1,
\end{equation}
and the $n_{\eta=\pm}(t)$ in Eq.~\eqref{liou3a} are expressed as
\begin{equation}\label{nplusmi}
n_\eta(t) =  P_\eta(t)+P_2(t). 
\end{equation}

\begin{figure}
\begin{center}
\includegraphics[width=0.48\textwidth]{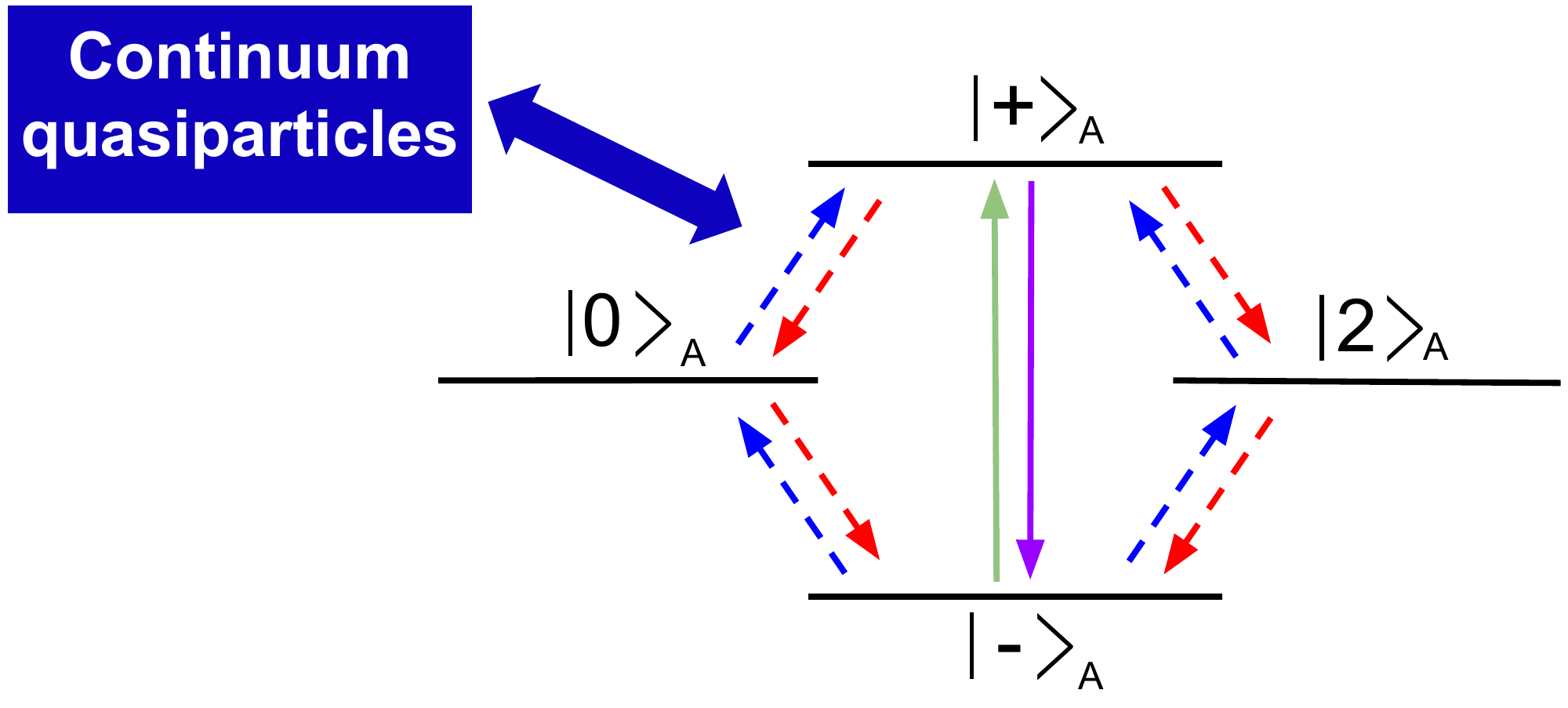}
\end{center}
\caption{\label{fig2}  Schematic illustration of the 
rate equation dynamics (see text). 
Direct transitions (solid arrows) connect the Andreev
level ground state, $|-\rangle_A$, to the excited 
state $|+\rangle_A$.
Transitions to the two degenerate odd-parity states (dashed arrows),
 $|0\rangle_A$ and $|2\rangle_A$,
are mediated through quasiparticle continuum states with energy
$|E|\ge \Delta$, which we indicate by a blue box. }
\end{figure}

We now observe that the off-diagonal components of $\rho_A(t)$ decouple 
from the equations for the diagonal part in
Eq.~\eqref{rhoAeq}; we briefly discuss the dynamics in the off-diagonal 
sector in Sec.~\ref{sec3b}.  The diagonal part of Eq.~\eqref{rhoAeq} 
determines the dynamics of the Andreev state occupation probabilities,
where we find
\begin{eqnarray} \label{liou3b}
\dot P_\eta &=& -\Gamma_{\eta,-\eta} P_\eta+ 
\Gamma_{-\eta,\eta} P_{-\eta} \\
\nonumber &-& \sum_p \Bigl[ n_p \left( \Gamma_{p,-\eta}P_\eta
-\Gamma_{p,\eta}P_0\right)\\ 
\nonumber
&& + (1-n_p) \left( \Gamma_{\eta,p} P_\eta - 
\Gamma_{-\eta,p}P_2\right)\Bigr].
\end{eqnarray} 
and
\begin{eqnarray}\label{liou3b1}
\dot P_0  &=& -\sum_{p,\eta=\pm} 
\left[ \Gamma_{p,\eta} n_p P_0 -\Gamma_{\eta,p}
(1-n_p) P_\eta \right], \\ \nonumber
\dot P_2 &=& -\sum_{p,\eta} 
\left[ \Gamma_{\eta,p}(1- n_p) P_2 - \Gamma_{p,-\eta}
n_p P_\eta \right].
\end{eqnarray}
Together with Eq.~\eqref{liou3a}, we thereby arrive at a set of 
coupled nonlinear equations determining the time-dependent 
continuum distribution function, $\{n_p(t) \}$,
and the Andreev level probabilities, $P_{\pm,0,2}(t)$. 
Importantly, despite of the approximations involved in their 
derivation, these coupled equations automatically satisfy
the  normalization condition \eqref{normP}.
The resulting Andreev bound state population dynamics 
is schematically illustrated in Fig.~\ref{fig2}. 
The rates $\Gamma_{\eta,-\eta}$  [Eq.~\eqref{directrate}]
connect the even-parity Andreev states $|\eta=\pm\rangle_A$, 
without involving continuum quasiparticles. However,
processes that populate or depopulate odd-parity Andreev states 
sensitively depend on the continuum distribution function $\{n_p(t)\}$. 

\subsection{Off-diagonal part of Andreev density matrix}\label{sec3b}

Within our approach, off-diagonal components of the Andreev 
density matrix $\rho_A(t)$ decouple from the diagonal ones
and obey their own set of dynamical equations.  
Although later on, we restrict ourselves to 
diagonal initial density matrices, where  
off-diagonal components do not appear at all, 
future experiments may test the corresponding quantum-coherent 
correlations in the Andreev sector.  Anticipating such 
experiments, which would probe an NMR-like quantum dynamics 
within the Andreev sector after careful choice of the initial conditions,
 we provide the relevant equations now.
Taking into account Hermiticity of the Andreev density matrix $\rho_A(t)$,
we find from Eq.~\eqref{rhoAeq} that 
$\rho_{0,2}(t)\equiv {}_A\langle 0|\rho_A(t)|2\rangle_A=
\rho_{0,2}(0)$, while the dynamics of the even-parity
matrix elements is determined by
\begin{eqnarray}
\partial_t \rho_{+,-} (t) &=&  - 
\frac12 \sum_\eta \Bigl[ \Gamma_{\eta,-\eta}+ \sum_{p}
\Bigl\{ n_p(t) \Gamma_{p,\eta}  \nonumber \\
&+& [1-n_p(t)]\Gamma_{\eta,p}\Bigr\}
\Bigr] \rho_{+,-}(t).
\label{offd1}
\end{eqnarray}
The dynamics of matrix elements connecting states with different parity
follows from
\begin{eqnarray} \nonumber 
\partial_t \rho_{\eta,0} &=& - \frac12  \Bigl \{
\Gamma_{\eta,-\eta} +\sum_p \Bigr[ n_p \left(
2\Gamma_{p,-\eta}+\Gamma_{p,\eta}\right)  \\ 
\label{offd2}
&+& (1-n_p) \Gamma_{\eta,p} \Bigr]\Bigr\} \rho_{\eta,0}  \\ \nonumber
&-& \eta\sum_p (1-n_p) \Gamma_{-\eta,p} \rho_{2, -\eta},
\end{eqnarray}
and 
\begin{eqnarray}\nonumber
\partial_t \rho_{2,-\eta} &=& - \frac12  \Bigl\{
\Gamma_{-\eta,\eta} +\sum_p \Bigl[ (1-n_p) \left(
2\Gamma_{-\eta,p}+\Gamma_{\eta,p}\right)  \\ 
\label{offd3} 
&+& n_p \Gamma_{p,\eta} \Bigr]\Bigr\} \rho_{2,-\eta} 
- \eta\sum_p n_p \Gamma_{p,-\eta} \rho_{\eta,0}.
\end{eqnarray}
In the remainder of the present work, however, we assume that
the initial state (at $t=0$) is diagonal.  In that case, the
decoupled off-diagonal density matrix elements remain zero 
during the entire time evolution.

\subsection{Steady-state distribution of quasiparticles} \label{sec3c}

We now proceed under the assumption that the initial Andreev density matrix,
$\rho_A(0)$, is diagonal in the basis $\{|\pm\rangle_A,|0\rangle_A,
|2\rangle_A\}$.  
In the long-time limit, the system will then reach a time-independent 
steady-state distribution, which is fully characterized by 
the probabilities $P_{\pm,0,2}$ 
together with the continuum quasiparticle distribution function $\{n_p\}$. 
In order to determine these quantities,
we first observe that $P_2=P_0$ due to the spin degeneracy
of the two odd-parity states.  
Moreover, by using the normalization condition in Eq.~\eqref{normP},
$P_0$ can be expressed in terms of $P_\pm$ alone,
\begin{equation}\label{p0p2}
P_0=P_2= \frac12 \left( 1- P_+-P_-\right).
\end{equation}
For the Andreev level occupations, we thus find 
\begin{equation}\label{neta1}
n_+ = 1-n_-= \frac12 \left(1+P_+ - P_{-} \right) ,
\end{equation}
and the steady-state version of Eq.~\eqref{liou3a} yields
\begin{equation}\label{steady1}
  0 =  - \sum_\eta \left[ \Gamma_{p,\eta} (1- n_{\eta}) n_p - 
\Gamma_{\eta,p} (1-n_p) n_\eta\right] - \frac{n_p^{}-n^{(0)}_p}{\tau_{\rm qp}} ,
\end{equation}
where we added a phenomenological relaxation term for continuum
quasiparticles describing, for instance, the effect of phonons \cite{olivares}. 
According to the estimates detailed in Ref.~\cite{olivares} for SACs made of aluminum, 
we expect $\tau_{\rm qp} \Delta\approx 10^4$ in the phonon-dominated regime (given by 
$E_A<\Delta-\Omega$, see below).  For simplicity, we here assume an energy-independent 
relaxation time, $\tau_{\rm qp}$, and a Fermi distribution function for the noninteracting 
continuum quasiparticles, 
\begin{equation}\label{Fermifunction}
n_{p=(E,s)}^{(0)}= \frac{1}{e^{E/T_{\rm qp}} +1},
\end{equation}
where the temperature $T_{\rm qp}$ may differ from the 
temperature $T_{\rm env}$ governing environmental phase fluctuations. 
We mention in passing that the theory in Ref.~\cite{olivares} corresponds
to the fast equilibration case with $\Gamma_{p,\eta} \tau_{\rm qp}\ll 1$.
Taking into account Eq.~\eqref{p0p2}, the rate equation \eqref{liou3b}
then yields the steady-state relation
\begin{eqnarray} \nonumber
0 &=& -\Gamma_{\eta,-\eta} P_\eta+ \Gamma_{-\eta,\eta} P_{-\eta} 
-\sum_p \Bigl[ n_p \Bigl( \Gamma_{p,-\eta}P_\eta \\ \label{steady2}
 & -& \Gamma_{p,\eta}P_0\Bigr) + (1-n_p)
 \left( \Gamma_{\eta,p} P_\eta - \Gamma_{-\eta,p}P_0 \right) \Bigr],
\end{eqnarray}
and Eq.~(\ref{liou3b1}) is automatically fulfilled. 

It is now a simple matter to solve Eq.~\eqref{steady1} for the
continuum quasiparticle distribution function,
\begin{equation}\label{steady1b}
n_p = \frac{\tilde\Gamma_p^{(-)}} {\tilde\Gamma_p^{(-)}+
\tilde\Gamma_p^{(+)} },
\end{equation}
which is thereby expressed by the $P_\pm$-dependent effective rates 
\begin{eqnarray} \label{tildegamma}
\tilde\Gamma_p^{(-)} &=& 
\sum_\eta \Gamma_{\eta,p} n_\eta + \frac{n_p^{(0)}}{\tau_{\rm qp}},
\\ \nonumber
\tilde \Gamma_p^{(+)} &=& 
\sum_\eta \Gamma_{p,\eta} (1-n_\eta) + \frac{1-n_p^{(0)}}{\tau_{\rm qp}}.
\end{eqnarray}
To obtain the Andreev level probabilities $P_\pm$, 
we then insert Eq.~\eqref{steady1b} back into Eq.~\eqref{steady2}.  
After some algebra, we arrive at two coupled \textit{nonlinear} equations,
\begin{eqnarray}\nonumber
&& \left( \begin{array}{cc} \Gamma_{+,-}+2 G_-  + G_+ &
G_+-\Gamma_{-,+}\\ G_--\Gamma_{+,-} & 
\Gamma_{-,+} +2G_++G_- \end{array}\right)
\left( \begin{array}{c} P_+ \\ P_- \end{array}\right) \\ 
\label{nonlin}
&& \quad \quad \quad = \left( \begin{array}{c} G_+ \\ G_- \end{array}\right),
\end{eqnarray} 
with the auxiliary functions 
\begin{eqnarray} \nonumber
&& G_{\eta=\pm}(P_+,P_-) = \frac{\nu_0}{2} \sum_{s=1}^4 
\int_{|E|\ge \Delta} dE \frac{|E|}{\sqrt{E^2-\Delta^2}} \\
&& \quad \quad \times \left[ 
 \Gamma_{p,\eta} n_p+
\Gamma_{\eta,p} (1-n_p) \right],
\end{eqnarray}
where $p=(E,s)$ and $\nu_0= L/(\pi\hbar v_F)$ is the normal density of states at the
Fermi level.  The nonlinear system in Eq.~(\ref{nonlin}) can be solved
by numerical iteration, where a relative accuracy of $10^{-6}$ was
ensured by using a Newton-Raphson algorithm.  This is necessary
because the continuum quasiparticle distribution $\{n_p\}$, which follows 
by virtue of Eq.~\eqref{steady1b} from the self-consistent solution for $P_+$ 
and $P_-$, strongly responds even to tiny changes in the $P_\pm$.

Below, it will be useful to consider the rate $\Gamma_{\rm in}$ for
transitions from the even-parity to the odd-parity sector (i.e.,
$P_0$ increases), as well as the escape rate, $\Gamma_{\rm out}$,
out of the odd-parity state (i.e., $P_0$ decreases).
Assuming an equilibrium quasiparticle distribution function $\{n_p\}$, 
those rates were considered in Ref.~\cite{olivares}. 
Here, by employing the self-consistent continuum quasiparticle distribution 
function, both rates can be read off from Eq.~\eqref{liou3b1},
\begin{equation}\label{inoutrate}
\Gamma_{\rm in} = \sum_{p, \eta} \Gamma_{\eta,p} (1 - n_p),\quad
\Gamma_{\rm out} = \sum_{p, \eta} \Gamma_{p, \eta} n_p.
\end{equation}
As observable of primary interest, we will 
discuss the quasiparticle current $I_{\rm qp}$, 
which follows with our self-consistent solution for $\{n_p\}$
by using standard scattering theory expressions. We summarize
these for the convenience of the reader in App.~\ref{appb}.

\subsection{Perfect transparency}\label{sec3d}

As application of our theory, we will discuss a charge imbalance 
effect in Sec.~\ref{sec4}.   This discussion is simplified 
when considering a SAC with perfect transparency, 
${\cal T}=1$.   
We specify the explicit form of the Andreev and continuum state wave 
functions for ${\cal T}=1$ in App.~\ref{appc}.  
The Andreev bound state energies, $\eta E_A$ with $\eta=\pm$, then follow from
$E_A(\varphi_0)= \Delta|\cos(\varphi_0/2)|$, see Eq.~\eqref{absen}, 
and for $\varphi_0\to \pi$, the Andreev levels tend to zero energy. 
Moreover, Eq.~\eqref{directrate} shows that transition rates between 
different Andreev states vanish for ${\cal T}=1$, i.e., 
$\Gamma_{\eta,-\eta}=0$.  

We show in App.~\ref{appc} that for given energy $E$ with $|E|\ge \Delta$, there are 
two decoupled types of scattering states $\Psi_{p=(E,s)}$, namely $s=\{1,4\}$ and
 $s=\{2,3\}$.  Those channels correspond to a net charge transfer across the weak link 
in \textit{opposite} directions.  Indeed, charge flows from the left to the 
right side for $s=\{1,4\}$, 
but from the right to the left when $s=\{2,3\}$, as is
directly seen from the definition of the scattering states, see 
Eqs.~\eqref{scattstate} and \eqref{scattsta}.  This also implies that the 
supercurrent matrix elements between Andreev and continuum states, ${\cal I}_{p,\eta}$, 
are nonzero only when $\eta =-{\rm sgn}(\pi-\varphi_0)$ for $s=\{1,4\}$, and $\eta=+{\rm sgn}
(\pi-\varphi_0)$ for $s=\{2,3\}$.  In what follows, we take the phase difference across the contact
as $0\le \varphi_0\le \pi$.  

With $\eta_E= {\rm sgn}(E)$,
some algebra then yields from Eq.~\eqref{rate} the 
transition rates \cite{olivares,foot3}
\begin{eqnarray}\label{olres}
\Gamma_{p=(E,s),\eta} &=& \frac{2\pi}{\hbar}  \frac{1}{4\pi\nu_0}
\frac{(E^2-\Delta^2)\sqrt{\Delta^2-E_A^2}} {|E| \omega_{\eta\eta_E}}  
\\ \nonumber &\times& \left[\delta_{\eta,-} (\delta_{s,1}+\delta_{s,4})
+ \delta_{\eta,+} (\delta_{s,2}+\delta_{s,3})\right] \\ \nonumber 
&\times& \left[ \delta_{\eta_E,+}+n_B\left(\omega_{\eta \eta_E}\right) 
\right] J(\omega_{\eta\eta_E}), \\ \nonumber
\omega_{\eta\eta_E=\pm} &=& |E|\mp E_A \ge 0, 
\end{eqnarray}
and similarly for $\Gamma_{\eta,p}$. 
In a transparent SAC, Eq.~(\ref{olres}) 
thus only allows for transitions between Andreev and continuum current 
states propagating in the same direction, which in turn causes 
the charge imbalance effect.  
Since the matrix elements in Eq.~(\ref{olres}) are identical 
for $s=\{1,4\}$ (and likewise for $s=\{2,3\}$),
the steady-state distribution function $n_{p=(E,s)}$ 
for continuum quasiparticles corresponds to a single
distribution function for left-movers, $n_L(E)$, and one for
right-movers, $n_R(E)$, respectively, 
\begin{eqnarray}\label{functio}
n_{(E,s=1)} &=&  n_{(E,s=4)}= n_R(E),\\ \nonumber
n_{(E,s=2)} &=&  n_{(E,s=3)}= n_L(E).
\end{eqnarray}
For $n_R(E)\ne n_L(E)$,  continuum quasiparticles are driven out of 
equilibrium.  For given steady-state Andreev occupation probabilities $P_\pm$,
the distribution functions in Eq.~(\ref{functio}) 
follow from Eqs.~\eqref{steady1b} and \eqref{tildegamma}, taken
with the above ${\cal T}=1$ rates.

\section{Charge imbalance effect}
\label{sec4}

We now discuss a charge imbalance effect which is predicted to be 
observable in high-transparency SACs.  We shall discuss this effect for a 
perfectly transmitting SAC, ${\cal T}=1$, and by assuming $\varphi_0\in
[0,\pi]$; for $\varphi_0\in (\pi,2\pi)$, the sign of the induced
quasiparticle current discussed below is reversed.
Noting that our theory allows for arbitrary $0<{\cal T}\le 1$,   
we find very similar results also for (not too small) ${\cal T}<1$ 
and observables taken as function of $E_A/\Delta$.
We then put ${\cal T}=1$ from now on.

Using Eqs.~\eqref{olres} and \eqref{qpcur}, 
the quasiparticle current flowing through the SAC is given by
\begin{equation}\label{chas}
I_{\rm qp} = \frac{e}{\pi\hbar}\int_{|E|\ge \Delta} dE \ j_{\rm qp}(E) 
\left[ n_R(E)-n_L(E) \right],
\end{equation}
with the energy-resolved dimensionless quasiparticle current ($|E|\ge \Delta$),
\begin{equation}
j_{\rm qp}(E) = \frac{|E|\sqrt{E^2-\Delta^2}}{ E^2-E_A^2 }, 
\end{equation}
and the self-consistent distribution functions 
$n_{L,R}(E)$ in Eq.~\eqref{functio}.
Evidently, if a charge imbalance is present, $n_L(E)\ne n_R(E)$, 
one generally expects a finite quasiparticle current $I_{\rm qp}$
from Eq.~\eqref{chas}. We also define the total accumulated 
quasiparticle charge,
\begin{equation}\label{qpcharge}
Q_{\rm qp}= e\nu_0 \int_{|E|\ge \Delta} dE \frac{|E|}{\sqrt{E^2-\Delta^2}} 
\left[ n_R(E)-n_L(E) \right].
\end{equation} 
Since the density of states $\nu_0\propto  L/\xi_0$,
where $L$ is the channel length and 
$\xi_0=\hbar v_F/\Delta$ the BCS coherence length,
$Q_{\rm qp}$ tends to vanish for 
a very short channel, $L/\xi_0 \to 0$,
while the induced quasiparticle current remains finite
in that limit.

\begin{figure}
\begin{center}
\includegraphics[width=0.45\textwidth]{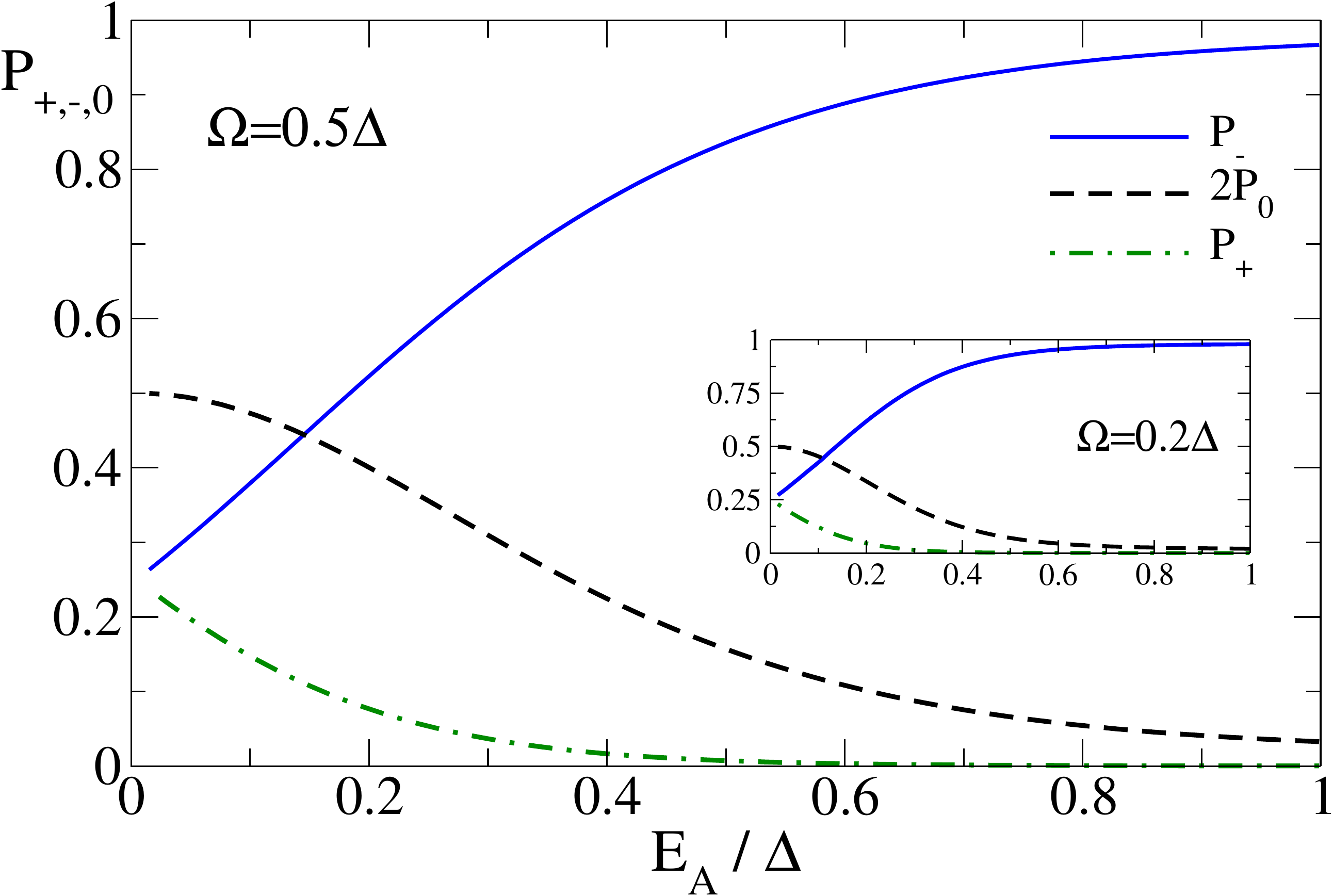}
\end{center}
\caption{\label{fig3} 
Self-consistent solution of Eq.~\eqref{nonlin} 
for the steady-state Andreev level occupation probabilities in a transparent
SAC, ${\cal T}=1$. Here, $P_{\rm \pm}$ refers to the even-parity Andreev levels,
with $|-\rangle_A$ being the ground state, and $P_{0}=P_2$ to
the degenerate pair of odd-parity states.  These results have been
obtained for plasma frequency $\Omega=0.5\Delta$, 
quasiparticle-photon coupling $\lambda=\sqrt{E_C/4\Omega}=
0.1$,  environmental temperature $T_{\rm env}=0.2\Delta$,
quasiparticle temperature $T_{\rm qp}=0.2\Delta$, 
channel length $L=\xi_0$, Ohmic damping constant 
$\eta_d=0.01\Delta$, and
$\tau_{\rm qp}\Delta=10^5$ (weak quasiparticle relaxation). 
The inset shows the case $\Omega=0.2\Delta$, 
$T_{\rm env}=0.5\Delta$, and 
$T_{\rm qp}=0.01\Delta$, where all other parameters are as in the main
panel.}
\end{figure}

Let us first address the steady-state Andreev populations, 
$P_{\pm,0,2}$, where $P_+$ and $P_-$ follow from the 
self-consistent solution of Eq.~\eqref{nonlin}, and the 
occupation probability
of the degenerate odd-parity state, $P_0=P_2$, is then given by 
Eq.~(\ref{p0p2}).  Representative results for $P_{\pm,0}$ vs $E_A/\Delta$ 
are shown for experimentally 
relevant parameters 
in Fig.~\ref{fig3}.  
Since $E_A=E_A(\varphi_0)$, see Eq.~\eqref{absen}, Fig.~\ref{fig3}
essentially shows the phase dependence of the Andreev state probabilities
for $\varphi_0\in [0,\pi]$.
The charge imbalance turns out to be absent in the strong relaxation regime
$\tau_{\rm qp}\Delta<1$ (see also below), where our theory reduces to the 
approach of Ref.~\cite{olivares} and thus self-consistency plays no role.
We therefore focus on the weak relaxation regime $\tau_{\rm qp}\Delta\gg 1$
in this section.  The main panel in Fig.~\ref{fig3} is 
for $T_{\rm env}=T_{\rm qp}$, while the inset studies a case 
where $T_{\rm env}$ substantially exceeds $T_{\rm qp}$.   
${}~$From Fig.~\ref{fig3}, we can distinguish two 
qualitatively different regimes, 
$E_A>\Delta-\Omega$ (with $P_0\to 0$) and $E_A<\Delta-\Omega$ (with
$P_0\ne 0$), respectively.
For $E_A>\Delta-\Omega$, the system remains quite close to the ground state, 
$|-\rangle_A$, since environmental photons can rapidly excite 
quasiparticles from an odd-parity state into the continuum.  
On the other hand, for $E_A<\Delta-\Omega$, the frequency $\Omega$ is too 
low to achieve such a transition.

\begin{figure}
\begin{center}
\includegraphics[width=0.45\textwidth]{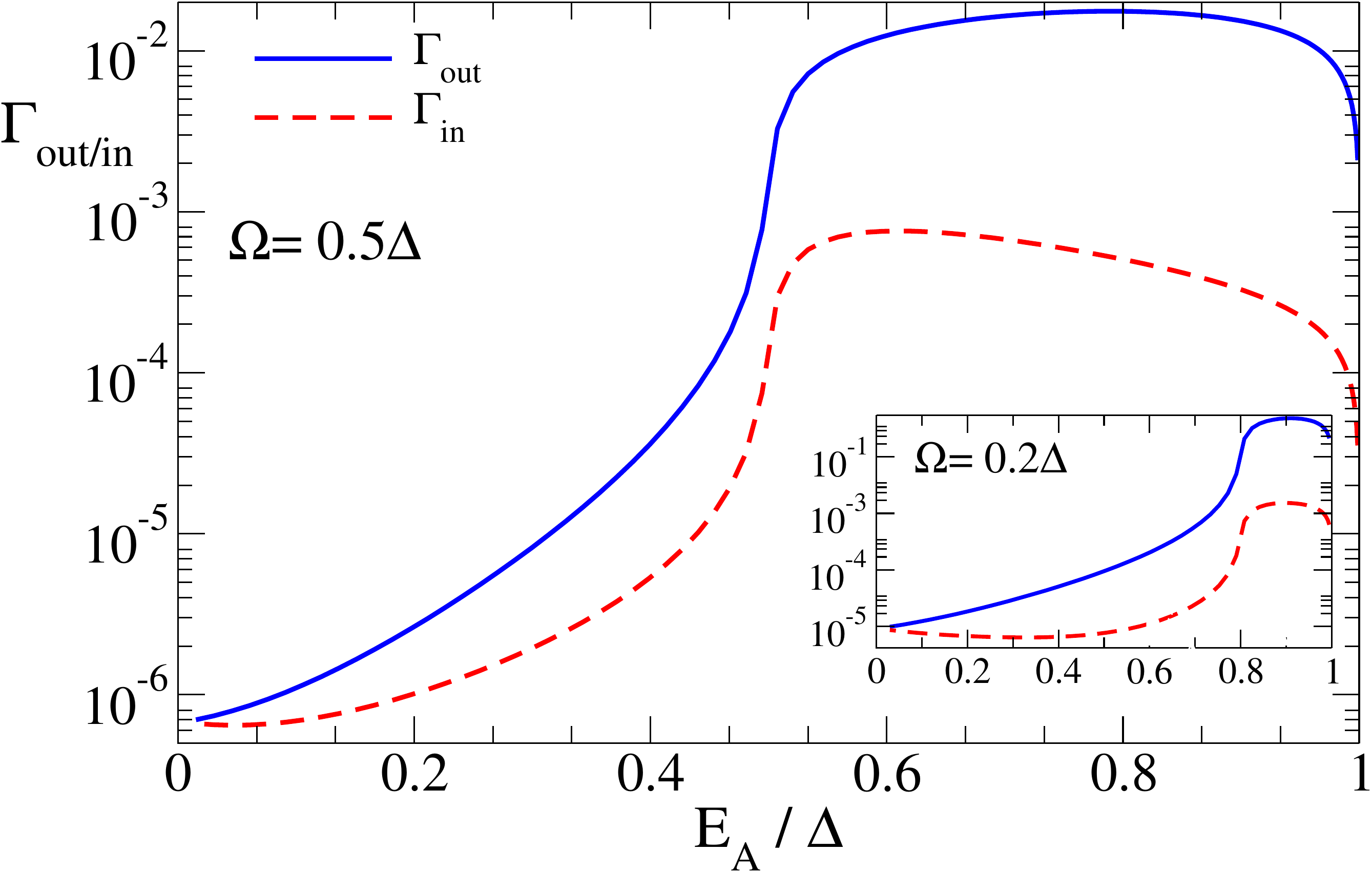}
\end{center}
\caption{\label{fig4} 
Transition rates $\Gamma_{\rm in}$ and $\Gamma_{\rm out}$ (in units 
of $\Delta/\hbar$) 
vs $E_A/\Delta$ on a semi-logarithmic scale.
$\Gamma_{\rm in}$ describes the rate for entering the odd-parity sector,
and $\Gamma_{\rm out}$ is the decay rate of odd-parity states.
Parameters are as in Fig.~\ref{fig3}. The inset shows the rates for 
parameters as in the inset of Fig.~\ref{fig3}.}
\end{figure}

The corresponding rates $\Gamma_{\rm in}$  and $\Gamma_{\rm out}$, see
Eq.~\eqref{inoutrate}, for populating and depopulating the odd parity states, 
resp., are shown in Fig.~\ref{fig4}, again as a function of $E_A/\Delta$. 
We observe now more clearly
that $E_A>\Delta-\Omega$ and
$E_A<\Delta-\Omega$ correspond to qualitatively different physical regimes.
For $E_A\approx \Delta-\Omega$, the rates increase over 
several orders in magnitude with very small
$\varphi_0$ variation, and one enters a regime where the odd-parity state
quickly decays.  This regime, $E_A>\Delta-\Omega$, has been termed
``fast relaxation regime'' in Refs.~\cite{zgirski,olivares}.

\begin{figure}
\begin{center}
\includegraphics[width=0.45\textwidth]{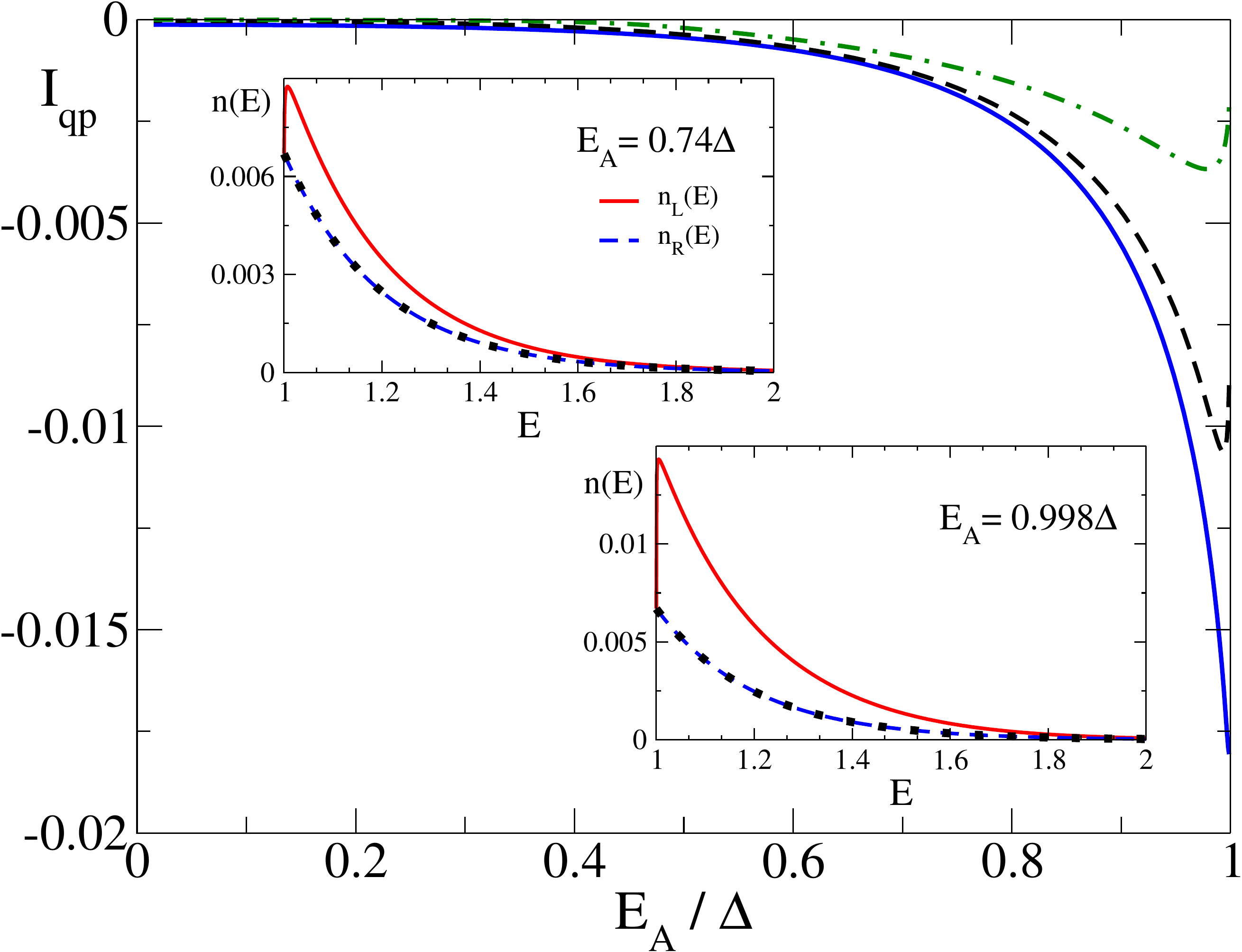}
\end{center}
\caption{\label{fig5} 
Main panel: Induced quasiparticle current $I_{\rm qp}$ (in units
of $e \Delta/\hbar$) vs $E_A/\Delta$ for varying 
$\tau_{\rm qp}\Delta = 10^5, 10^4$ and $10^3$ from bottom to top; other
parameters are as in the main panel of Fig.~\ref{fig3}.  
Insets:
Continuum quasiparticle distributions
$n_{L,R}(E)$ vs $E/\Delta$ for two $E_A/\Delta$ values
and $\tau_{\rm qp}\Delta = 10^5$.
For $E<0$, the distribution functions follow by using the 
electron-hole symmetry relation $n_R(-E)=1-n_L(E)$.
Dotted curves indicate the corresponding 
equilibrium Fermi distributions.}
\end{figure}

Next, in Fig.~\ref{fig5} we show the induced quasiparticle 
current $I_{\rm qp}$, see Eq.~\eqref{chas}, vs $E_A$ for $\Omega=0.5\Delta$. 
This quantity clearly demonstrates
that there is a significant charge imbalance effect throughout
the regime $E_A>\Delta-\Omega$, but not for $E_A<\Delta-\Omega$. 
The induced current gets reduced as the quasiparticle relaxation rate
$1/\tau_{\rm qp}$ increases,
 and is only significant for $\tau_{\rm qp} \Delta \gg 1$, which
is the typical regime for SACs made of aluminum~\cite{zgirski}. 
Further insight on the generated charge imbalance is 
obtained by analyzing the distribution
functions $n_{R,L}(E)$ for right- and left-moving quasiparticles, 
see Eq.~\eqref{functio}.  As illustrated by the insets in 
Fig.~\ref{fig5}, the generated imbalance is maximal
 for $E_A \to \Delta$ (i.e., for $\varphi_0\to 0$, where 
the supercurrent $\langle \hat I_S\rangle$ can be vanishingly small), and
becomes smaller as $E_A$ decreases. For the present case
with $T_{\rm qp}=T_{\rm env}$, 
the smaller $n(E)$ curve (i.e., the $n_R$ component for $E>0$,
 and $1-n_L$ for $E<0$) coincides
with the Fermi distribution at the corresponding temperature, 
indicated by dotted curves in the insets.
Noting that the Josephson current for a fully transparent SAC is of 
order $\langle \hat I_S\rangle =\langle \hat I_A\rangle
\approx e\Delta/ \hbar$, the induced quasiparticle current
is a few percent of this value for the parameters in Fig.~\ref{fig5}.  
In the ring geometry of Fig.~\ref{fig1}, the quasiparticle current $I_{\rm qp}$ flows
in opposite direction to the Josephson current $\langle \hat I_S\rangle$, which 
can be rationalized as follows.
The rate from $|+\rangle_A$ to the left-moving $s=2$ 
continuum states with $E>0$ --- carrying negative current --- is much larger 
than the one from $|-\rangle_A$ to the $(E>0,s=1)$
states carrying positive current,  due to
the much shorter distance in energy.  This difference is able
to outweigh the fact that $P_+<P_-$ favors the same sign of $I_{\rm qp}$
and $\langle \hat I_S\rangle$.

\begin{figure}
\begin{center}
\includegraphics[width=0.45\textwidth]{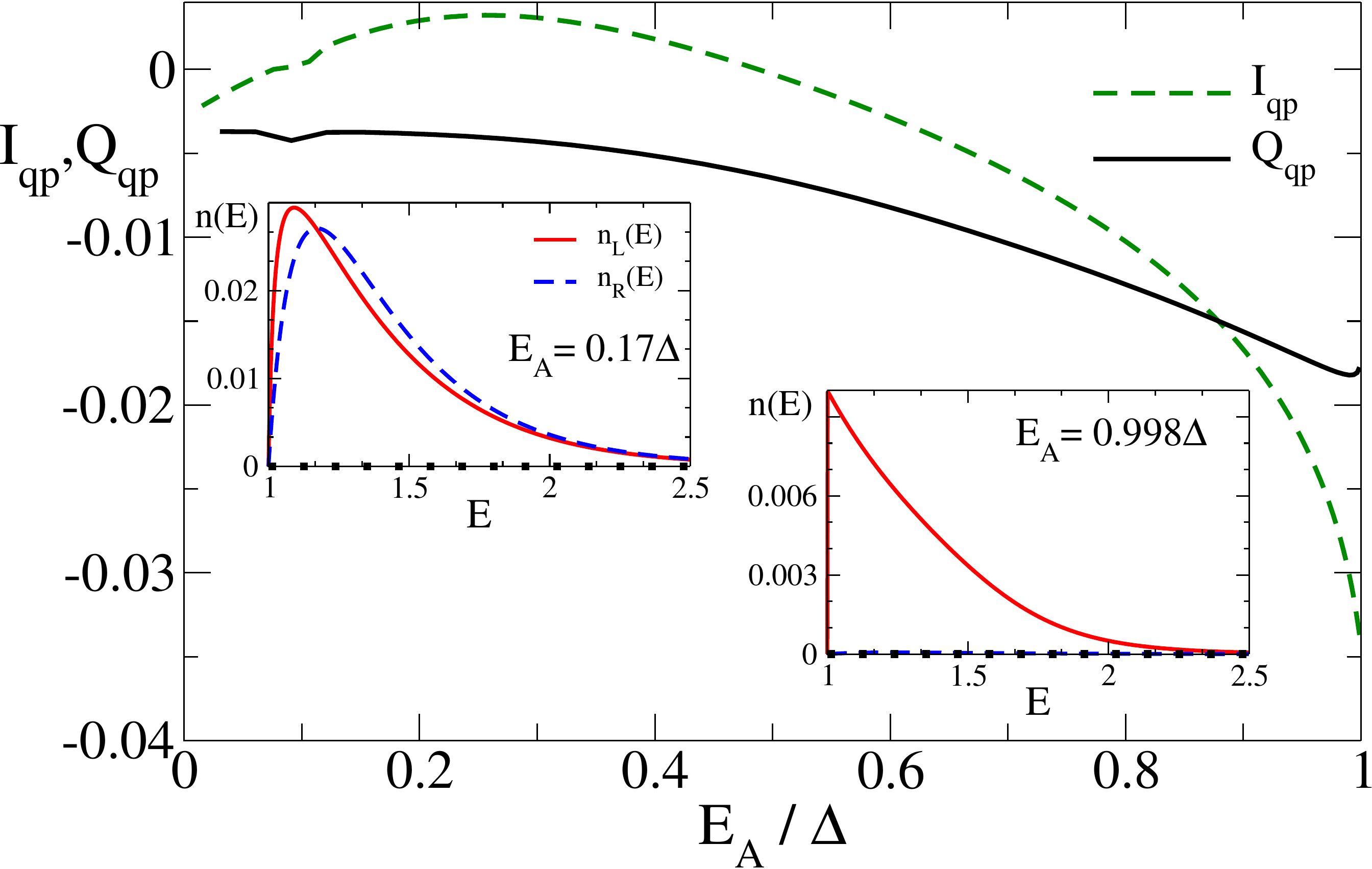}
\end{center}
\caption{\label{fig6}  
Main panel: Quasiparticle current $I_{\rm qp}$ (in $e\Delta/\hbar)$ 
and accumulated charge $Q_{\rm qp}$ (in units of e) vs $E_A/\Delta$ 
for the parameters in the inset of Fig.~\ref{fig3}, i.e., $\Omega=0.2\Delta,
T_{\rm env}=0.5\Delta$ and $T_{\rm qp}= 0.01\Delta$.
The insets show the continuum
quasiparticle distributions, $n_{R,L}(E)$, for two different $E_A$ values.
 In contrast to the case studied in Fig.~\ref{fig5},
 the induced quasiparticle current is now 
significant for the whole $E_A$ range, and exhibits a sign change for
 $E_A \simeq \Omega$.  
}
\end{figure}

The parameters considered up to now were inspired by those 
realized in available experimental reports \cite{zgirski,olivares}.
However, as we show next, it is also interesting to consider 
an alternative scenario, where the temperature of the environmental modes 
is so high to put them into a classical regime, $T_{\rm env}\gg \Omega$.
Experimentally, such a situation can be realized by
replacing the electromagnetic environment by an external microwave 
radiation source of frequency $\Omega$.  We here consider the 
case $\Omega=0.2\Delta$, with $T_{\rm env}=0.5\Delta=2.5\Omega$.
For the quasiparticle temperature $T_{\rm qp}$, we take 
$T_{\rm qp}=0.01\Delta$,
significantly smaller than $T_{\rm env}$. 
 The Andreev state populations for this case were shown in the inset
of Fig.~\ref{fig3}, and the corresponding $\Gamma_{\rm in/out}$ 
rates in the inset of Fig.~\ref{fig4}.
Again fast and slow relaxation regimes
(in the parlance of Ref.~\cite{zgirski}) can be identified for 
$E_A >\Delta-\Omega$ and
$E_A < \Delta -\Omega$, respectively. However, in this
 case, the generated quasiparticle
populations differ more strongly from the Fermi
 distributions (see insets of Fig.~\ref{fig6}),
and a significant quasiparticle current is induced 
throughout the whole $E_A$ range. This is illustrated
in the main panel of Fig.~\ref{fig6}. It is also interesting 
to notice that the induced
quasiparticle current exhibits a sign change for $E_A \simeq \Omega$.  

\section{Concluding remarks}\label{sec5}

In this work, we have formulated and applied a theoretical framework
for the Andreev bound state population dynamics in single-channel 
superconducting weak links.  Taking into account phase fluctuations
by an electromagnetic environment,
we have developed a master equation approach for the quasiparticle
dynamics, capturing the rich interplay between Andreev states and
continuum states.  In particular, the role of odd-parity Andreev 
states and the need for a self-consistent treatment of the 
generated nonequilibrium continuum quasiparticle distribution has been 
emphasized.  As an application of our theory, we have 
demonstrated that the asymmetry in Andreev-continuum quasiparticle 
transition rates causes an intriguing charge transfer across the
weak link, reflected in a quasiparticle current. 
Using established experimental techniques,
this charge imbalance should be measurable in a SAC.
Our theory could be also applied for the study of the quantum coherent dynamics 
of this system, including the effect of parity mixing processes. 
This is of relevance for
the various proposals of using Andreev levels as qubits \cite{chtech,padurariu,zazunov2,desposito}. 
As another extension of our formalism, it would be very interesting
to study the Andreev- and Majorana bound state dynamics in topological
superconductor weak links, or to study the interaction-induced effects
(see also Ref.~\cite{remi})
on Andreev bound state dynamics when the constriction contains a quantum 
dot with sizeable charging energy, or couples to local phonon modes.

\textit{Note:} During the preparation of this manuscript, we became aware of
related work \cite{yuli1}, where the
charge imbalance effect described here has also been pointed out.  
Where there is overlap, our results match theirs.
However, in contrast to our work,  Ref.~\cite{yuli1} considers 
a monochromatic external microwave driving and the 
continuum quasiparticle distribution was not treated in a 
self-consistent manner.  

\acknowledgments
 We thank Yu.V. Nazarov for discussions.
This work has been supported by the DFG networks SFB-TR 12 and SPP 1666, 
 by the DFG grant No.~EG 96/9-1, and from the Spanish MINECO 
through project No.~FIS2011-26516.

\appendix
\section{Quasi-particle wave functions}\label{appa}

In this Appendix, we provide the wave functions, 
$\Psi_\nu(x)$, solving the stationary BdG equation, 
$H_0 \Psi_\nu=E_\nu \Psi_\nu$,
under the matching condition \eqref{matching} for 
time-independent phase difference, $\varphi(t)=\varphi_0$, 
with $0\le \varphi_0<2\pi$.

Andreev bound states, $\nu=\eta=\pm$, with energy $\eta 
E_A(\varphi_0)$, see Eq.~\eqref{absen}, have the wave function
\begin{eqnarray}\label{absstate}
\Psi_\eta(x) &=& \xi_0^{-1/2} e^{-\sqrt{\cal T}\sin(\varphi_0/2) |x|/\xi_0} \\ 
\nonumber &\times&  \left[ \Theta(-x) 
\left( \begin{array}{c}
A_\eta\tilde\psi_h \\ B_\eta\tilde\psi_e \end{array}\right)
 + \Theta(x) \left(\begin{array}{c} -\eta A_\eta \tilde\psi_e\\
 \eta B_\eta \tilde\psi_h \end{array}\right)\right], 
\end{eqnarray}
where $\Theta(x)$ the Heaviside step function.  We use the Nambu spinors
\begin{equation}\label{psieh}
\tilde\psi_{e,h} = \frac{ e^{\pm i\theta_\eta\sigma_z/2} }{\sqrt{2}}
\left(\begin{array}{c} 1 \\ \eta \end{array} \right),
\end{equation}
where $\cos\theta_\eta= E_A/\Delta$ with 
$\eta\sin\theta_\eta\ge 0$. We also define the parameters
\begin{eqnarray}\label{aeta}
A_\eta&=& \sqrt{{\cal N}_\eta}\sin(\varphi_0/2-\theta_\eta),\\ \nonumber
B_\eta &=& \sqrt{{\cal N}_\eta (1-{\cal T})}\sin(\varphi_0/2),\\ \nonumber 
{\cal N}_\eta &=&
\frac{\sqrt{\cal T}}{2\cos(\theta_\eta)
\sin(\varphi_0/2-\theta_\eta)}.
\end{eqnarray}
The Andreev bound states \eqref{absstate} satisfy the normalization 
condition $\int dx\  \Psi_\eta^\dagger(x) \cdot 
\Psi_{\eta^\prime}(x) = \delta_{\eta\eta^\prime}.$

Next we summarize the stationary solutions of the
 BdG equation in the continuum, $\Psi_{p=(E,s)}(x)$ with $|E|\ge \Delta$.  
Using $\eta_E= {\rm sgn}(E)=\pm$ and $\cosh\theta_E=|E|/\Delta$ 
(with $\theta_E\ge 0$), and denoting the wavenumber by
$k=\eta_E\sqrt{E^2-\Delta^2}/v_F$, we find
\begin{eqnarray} \nonumber
\Psi_{p} &=& \Psi_p^{(\rm in)}+ \Theta(-x)
\frac{e^{-ikx}}{\sqrt{2L}} \left( \begin{array}{c}
a\psi_h \\ b\psi_e \end{array}\right) \\ &+& \label{scattstate}
\Theta(x)\frac{e^{ikx}}{\sqrt{2L}} \left(\begin{array}{c} c\psi_e\\
 d  \psi_h \end{array}\right),
\end{eqnarray}
where the electron- and hole-type Nambu spinors $\psi_{e,h}$
 follow by analytic continuation of Eq.~\eqref{psieh},
\begin{equation}\label{nambu2}
\psi_{e,h}=\frac{e^{\pm \theta_E \sigma_z/2}}{\sqrt{2\cosh\theta_E}}
\left( \begin{array}{c} 1\\ \eta_E\end{array}\right).
\end{equation}
There are four different solutions ($s=1,2,3,4$), 
describing electron- or hole-type states 
incoming from the left or right side,
\begin{equation}\label{scattsta}
\Psi_p^{(\rm in)} = \Theta(-x) \frac{e^{ikx}}{\sqrt{2L}} \left(
\begin{array}{c} \psi_e \delta_{s,1}
\\  \psi_h\delta_{s,2} \end{array} \right) + 
\Theta(x) \frac{e^{-ikx}}{\sqrt{2L}} 
\left(\begin{array}{c} \psi_h \delta_{s,4} \\ 
\psi_e\delta_{s,3}\end{array}\right).
\end{equation}
With $Q= \sinh^2\theta_E + {\cal T}\sin^2(\varphi_0/2),$
the scattering amplitudes ($a,b,c,d$) appearing in Eq.~\eqref{scattstate} 
can be expressed in terms of four functions,
\begin{eqnarray}\nonumber
A(\theta,\varphi) &=& -\frac{i{\cal T}}{Q}
\sin(\varphi/2)\sinh(\theta-i\varphi/2),\\
B(\theta,\varphi) &=& \frac{\sqrt{1-{\cal T}}}{Q} \sinh^2\theta  ,\\ \nonumber
C(\theta,\varphi) &=&\frac{\sqrt{\cal T}}{Q}\sinh(\theta)
\sinh\left(\theta-i\varphi/2\right),\\ \nonumber
D(\theta,\varphi) &=& \frac{i\sqrt{(1-{\cal T}){\cal T}}}{Q}
\sin(\varphi/2)\sinh\theta,
\end{eqnarray}
such that for $s=1$,
\begin{equation}\label{as1}
\left(\begin{array}{c} a_1 \\ b_1 \\ c_1 \\ d_1 \end{array}
\right) = \left( \begin{array}{c} 
A(\theta_E,\varphi_0)\\
B(\theta_E,\varphi_0)\\
C(\theta_E,\varphi_0)\\
D(\theta_E,\varphi_0)\end{array}\right) 
\end{equation}
For the other three possible values of $s$, we find
\begin{eqnarray}\label{bs1}
\left(\begin{array}{c} a_2 \\ b_2 \\ c_2 \\ d_2 \end{array}
\right) &=& \left( \begin{array}{c} 
B(-\theta_E,\varphi_0)\\
A(-\theta_E,\varphi_0)\\
D(-\theta_E,\varphi_0)\\
C(-\theta_E,\varphi_0)\end{array}\right) ,
\\ \nonumber
\left(\begin{array}{c} 
a_3 \\ b_3 \\ c_3\\ d_3 \end{array}
\right) &=& \left( \begin{array}{c} 
-D(\theta_E,-\varphi_0)\\
C(\theta_E,-\varphi_0)\\
-B(\theta_E,-\varphi_0)\\
A(\theta_E,-\varphi_0)\end{array}\right) ,\\ \nonumber
\left(\begin{array}{c} 
a_4 \\ b_4 \\ c_4\\ d_4 \end{array}
\right) &=& \left( \begin{array}{c} 
C(-\theta_E,-\varphi_0)\\
-D(-\theta_E,-\varphi_0)\\
A(-\theta_E,-\varphi_0)\\
-B(-\theta_E,-\varphi_0)\end{array}\right) .
\end{eqnarray}
Notice that for all $s$, the relation $ab+cd=0$ is fulfilled.

\section{Supercurrent matrix elements}\label{appb}

In this appendix, we discuss the matrix elements necessary for the 
evaluation of the Josephson current operator, see Eq.~\eqref{joscur}. 
The matrix elements determining the pure Andreev contribution are
readily obtained and have been specified in Eq.~\eqref{iacur}.
We then address the matrix elements ${\cal I}_{\eta,p}$
entering $\hat I_{cA}$ in Eq.~\eqref{ica},
which describe the mixing of the Andreev bound state at energy $\eta E_A$,
with $\eta=\pm$, and the continuum
state with $p=(E,s)$, where $\eta_E={\rm sgn}(E)$ and $|E|\ge \Delta$.
The index $s$ describes the four types of scattering states, 
see App.~\ref{appa}.
{}From Eq.~\eqref{joscur}, we first need to determine the corresponding
matrix elements $A_{\eta,p}$ and $W_{\eta,p}$. After some algebra,
using the auxiliary quantities
\begin{eqnarray}
u &=& \frac{1}{\eta \sin \theta_\eta + i\eta_E \sinh\theta_E},
\\ \nonumber
z &=& \frac12 \left( e^{ (\theta_E+i\theta_\eta)/2 }
-\eta\eta_E e^{-(\theta_E+i\theta_\eta)/2} \right),
\end{eqnarray}
as well as the definitions in App.~\ref{appa}, we find
\begin{eqnarray} \label{wadef}
&& \left(\begin{array}{c} W_{\eta,p}/\Delta \\ 2 A_{\eta,p} \end{array}\right) 
=  \sqrt{\frac{\xi_0}{8L\cosh\theta_E}} \times \\  \nonumber&&  \Bigl\{ u^*
[ (c-\eta_E a)\eta A_\eta  +  (b+\eta_E d)B_\eta ] 
\left(\begin{array}{c} i\eta z \\ z^*\end{array}\right) \\ \nonumber
&& + u \left[ (\delta_{s,1}-\eta_E\delta_{s,4})A_\eta -
(\eta_E\delta_{s,2}+\delta_{s,3})  \eta B_\eta \right] 
\left(\begin{array}{c} i\eta z^* \\ z\end{array}\right)
\Bigr\}. 
\end{eqnarray} 
Equation~\eqref{joscur} then yields the 
current matrix elements ${\cal I}_{\eta,p}$.

We next show that matrix elements between continuum states 
vanish identically, ${\cal I}_{pp'}=0$.  In the limit $L\to \infty$,
only states with $E=E'$ can have a finite matrix element. 
Taking into account that the Nambu spinors \eqref{nambu2}
satisfy the relations $\bar\psi_{e,h}\sigma_y \psi_{e,h}=0$ and
$\bar\psi_{e,h}\sigma_z \psi_{e,h}=\pm \tanh\theta_E$,
one then finds $W_{pp'}=0$.  
Although the matrix elements $A_{pp'}$ are nonzero, they do not contribute
to ${\cal I}_{pp'}$ because they appear together with a factor $(E-E')=0$.
Transitions between continuum states can therefore not contribute
to the Josephson supercurrent operator $\hat I_S$. 

Finally, the continuum contribution to the dissipative
quasiparticle current, $I_{\rm qp}$, 
follows from the $\Psi_p$ in Eq.~\eqref{scattstate},
\begin{equation}
I_{\rm qp}= ev_F\sum_{p=(E,s)} n_p 
\bar{\Psi}_p \tau_z \Psi_p,
\end{equation}
where the Pauli matrix $\tau_z$ acts in left-right mover space,
see Sec.~\ref{sec2}.   Using the $s$-dependent 
scattering amplitudes $(a,b,c,d)$ in App.~\ref{appa},
we find
\begin{eqnarray}\label{qpcur}
&& I_{\rm qp} = \frac{e}{2\pi \hbar} \sum_{s=1}^4 \int_{|E|\ge \Delta}
\frac{|E|dE}{\sqrt{E^2-\Delta^2} } \\  &\times & n_{(E,s)}  \nonumber
\Bigl\{ (\delta_{s,1}+\delta_{s,2}) \left[\left|c_s\right|^2-\left|
d_s\right|^2\right]  \\ &+& \nonumber
(\delta_{s,3}+\delta_{s,4}) \left[\left|a_s\right|^2-\left|
b_s\right|^2\right] \Bigr\}.
\end{eqnarray}

\section{Perfect transparency}\label{appc}

Here we summarize the quasiparticle wave functions
for ideal contact transparency, ${\cal T}\to 1$.  
In the Andreev bound state wave functions, $\Psi_{\eta=\pm}(x)$ 
in Eq.~\eqref{absstate}, the coefficients 
 $A_\eta$ and $B_\eta$ now take the form ($0\le \varphi_0<2\pi$)
\begin{eqnarray}
A_\eta &=& \sqrt{\sin(\varphi_0/2)} \ \delta_{\eta,-{\rm sgn}(\pi-\varphi_0)}, \\  \nonumber
B_\eta &=& \sqrt{\sin(\varphi_0/2)} \ \delta_{\eta,{\rm sgn}(\pi-\varphi_0)}.
\end{eqnarray}
Turning to the continuum state wave functions 
$\Psi_{p=(E,s)}(x)$ in Eq.~\eqref{scattstate}, we need 
the scattering amplitudes $(a_s,b_s,c_s,d_s)$ for an 
incoming state of type $s=\{1,2,3,4\}$, which have been 
specified for arbitrary ${\cal T}$ in 
Eqs.~\eqref{as1} and \eqref{bs1}.
For ${\cal T}=1$, these results can be simplified to yield
\begin{eqnarray*}
\left(\begin{array}{c} a_1\\ b_1\\ c_1\\ d_1\end{array}\right) &=&
\left(\begin{array}{c} c_4\\ b_4\\ a_4\\ d_4\end{array}\right) =
\frac{1}{\sinh(\theta_E+i\varphi_0/2)} 
\left(\begin{array}{c} -i\sin(\varphi_0/2)\\ 0\\ \sinh\theta_E \\ 0
\end{array}\right), \\ 
\left(\begin{array}{c} a_2\\ b_2\\ c_2\\ d_2\end{array}\right) &=&
\left(\begin{array}{c} a_3\\ d_3\\ c_3\\ b_3\end{array}\right) =
\frac{1}{\sinh(\theta_E-i\varphi_0/2)} 
\left(\begin{array}{c} 0\\ i\sin(\varphi_0/2)\\ 0 \\ \sinh\theta_E
\end{array}\right).
\end{eqnarray*}


\begin{thebibliography}{12}
\bibitem{foot1}
Specifically, the Markov approximation amounts to replacing 
$\rho(t-\tau)\to \rho(t)$ and neglecting the last term (describing
correlations with the initial state) in Eq.~\eqref{liou1}.
This step is valid for temperatures above $\lambda^2/\Delta$.
\bibitem{foot2} 
Mathematically speaking, we here construct an asymptotic solution 
for the reduced density matrix of the form 
$\rho_0(\zeta t) + \zeta \rho_1(t)$, where
$\zeta$ is a small expansion parameter $\propto \lambda^2\ll 1$,
which reflects the weakness of the coupling $\lambda$.
The first term corresponds to Eq.~\eqref{factoriz} and
is ``slow'' on the timescale $1/\Delta$, while the second term
contains off-diagonal terms oscillating with frequencies 
$\agt (\Delta-E_A)$.  
This term is therefore ``fast'' and remains small.
\bibitem{foot3}
Note that for each scattering channel $s$, 
$|{\cal I}_{\eta,p=(E,s)}|^2$ in Eq.~\eqref{olres} is invariant under a 
particle-hole transformation, $(E,\eta E_A)\to (-E,-\eta E_A)$.
\end{thebibliography}
\end{document}